\documentclass[12pt,preprint]{aastex}
\newcommand{\beq}{\begin{equation}}
\newcommand{\beqa}{\begin{eqnarray}}
\newcommand{\eeq}{\end{equation}}
\newcommand{\eeqa}{\end{eqnarray}}
\newcommand{\simg}{\gtrsim}
\newcommand{\siml}{\lesssim}
\newcommand{\meszaros}{M${\acute {\rm e}}$sz${\acute {\rm a}}$ros}
\shorttitle{TeV-PeV Neutrinos from SGR 1806-20}
\shortauthors{Ioka, Razzaque, Kobayashi, \& \meszaros}
\begin{document}
\title{
TeV-PeV Neutrinos from Giant Flares of Magnetars and the Case of SGR 1806-20
}
\author{
Kunihito Ioka\footnote{
Physics Department and Center for Gravitational Wave Physics, 
104 Davey Laboratory, Pennsylvania State University, University Park,
PA 16802},
Soebur Razzaque\footnote{
Department of Astronomy and Astrophysics, 525 Davey Laboratory,
Pennsylvania State University, University Park, PA 16802},
Shiho Kobayashi$^{1,2}$,
and Peter \meszaros$^{1,2}$
}

\begin{abstract}
We estimate the high energy neutrino flux from the giant flare of SGR 1806-20
on December 27, 2004, which irradiated Earth with a gamma-ray flux  $\sim 10^4$ 
times larger than the most luminous gamma-ray bursts (GRBs) ever detected.
The Antarctic Cherenkov neutrino detector AMANDA
was on-line during the flare, 
and may either have detected high energy 
neutrinos for the first time from a cosmic point source, or put constraints 
on the flare mechanism of magnetars.
If TeV neutrinos are detected, one would expect also detectable
EeV cosmic rays and possibly TeV gamma-ray emission in coincidence.
\end{abstract}


\section{Introduction}

The giant flare of SGR 1806-20 on December 27, 2004 was the brightest 
cosmic transient to date, emitting gamma-rays for $\sim 0.1$ sec with 
flux $\sim 10$ erg s$^{-1}$ cm$^{-2}$ 
\citep{terasawa05,hurley05,mazets05,palmer05}.
If a comparable energy was emitted as high energy neutrinos, they should 
have been detected for the first time by current neutrino observatories 
such as AMANDA \citep{ahrens02}. 
In this Letter we calculate the expected high energy 
neutrino flux from giant flares in Soft Gamma-ray Repeaters (SGRs) such as 
SGR 1806-20, and argue that the neutrino flux is indeed either detectable, 
or its absence provides important constraints on the flare mechanism
(see \citet{zhang03} for high energy neutrinos from quiescent magnetars).

SGRs are a type of extreme X-ray pulsars, repeatedly emitting $\sim 0.1$ sec 
bursts of soft gamma-rays. Giant flares are more energetic events, which 
have been recorded from three of four known SGRs with the December 27 event 
being the third one. SGRs are most likely magnetars, i.e., highly magnetized 
neutron stars \citep{thompson95,thompson01}.  In this model giant flares 
result from a global magnetic rearrangement of the crust or even the entire 
interior \citep[e.g.,][]{ioka01}.

Giant flares have many similarities to cosmological gamma-ray bursts (GRBs).
Long duration GRBs are thought to arise from relativistic jets interacting 
with themselves for prompt GRBs, and subsequently with a circumburst medium 
for the longer wavelength afterglows \citep[e.g.,][]{zhang04}. In a similar 
manner, the detected radio afterglows of SGRs imply the presence of 
relativistic outflows \citep{frail99,cameron05,gaensler05}, and the huge 
flare luminosities also lead to relativistic fireballs 
\citep{huang98,thompson01,nakar05}. In particular, the minimum energy of the 
radio afterglow is larger than the kinetic energy of $e^{\pm}$ pairs that 
survive annihilation, implying the presence of baryons in fireballs like GRBs 
\citep{nakar05}. Relativistic baryons are likely to cause shocks, leading to 
Fermi accelerated protons, and to high energy neutrinos via $p\gamma$ 
\citep{waxman97} and $pp$ interactions \citep{paczynski94}. The accelerated 
protons in GRBs may also explain the observed ultrahigh energy cosmic rays 
\citep{waxman95,vietri95,ioka04}.

In contrast to GRBs, however, the baryon load in giant SGR flares is less 
constrained (see \S~\ref{sec:fire}), mainly because the flare spectrum 
may be thermal \citep{hurley05} or nonthermal 
\citep[][]{mazets05,palmer05}. Since the neutrino fluxes 
depend on the baryon load, neutrinos can be a probe of the baryon load as 
well as the flare mechanism. In \S~\ref{sec:fire} we discuss typical 
fireball models for SGR 1806-20 and in \S~\ref{sec:neutrino} we estimate 
the expected high energy neutrino fluxes for these models. The implications 
are discussed in \S~\ref{sec:dis}.

\section{Typical fireball models  for SGR 1806-20}\label{sec:fire}

The giant flare of SGR 1806-20 on December 27, 2004 radiated a gamma-ray 
energy $E_{\gamma} \sim 3 \times 10^{46} E_{\gamma,46.5}$ erg  during a
time of $t_{0} \sim 0.1 t_{0,-1}$ sec \citep{terasawa05}.\footnote{
We adopt $d=10d_{1}$ kpc for the distance to SGR 1806-20 though it is 
controversial \citep{cameron05,corbel04}.}
The total luminosity was $L_{0} \sim L_{\gamma}/\xi_{\gamma} \sim
3 \times 10^{47} L_{0,47.5}$ erg s$^{-1}$ 
for a conversion efficiency $\xi_{\gamma}$ of total energy into gamma-ray.
If such energy is released near a neutron star (radius $r_{0} \sim 10^{6} 
r_{0,6}$ cm), $e^{\pm}$ pair production creates an optically thick fireball 
with an initial temperature \citep{paczynski86,goodman86}
\beqa
T_{0}\sim \left(\frac{L_{0}}{4\pi r_{0}^{2} c a}\right)^{1/4}
\sim 300 L_{0,47.5}^{1/4} r_{0,6}^{-1/2} {\rm \ keV},
\eeqa
where $a=\pi^{2} k^{4}/15 \hbar^{3} c^{3} = 7.6 \times 10^{-15}$ 
erg cm$^{-3}$ K$^{-4}$ is the radiation density constant.
As the fireball expands under its own pressure,
the Lorentz factor increases as $\Gamma \propto r$
and the comoving temperature drops as $T \propto r^{-1}$.
The subsequent evolution depends on the baryon load
parametrized by a dimensionless entropy $\eta=L_{0}/\dot M c^{2}$ \citep{shemi90}.
If the fireball is baryon-rich, $\eta < \eta_{*}$, where
$\eta_{*}$ is a critical entropy \citep{meszaros00}
\beqa
\eta_{*} = \left(\frac{L_{0} \sigma_{T}}{4\pi m_{p} c^{3} r_{0}}\right)^{1/4}
\sim 100 L_{0,47.5}^{1/4} r_{0,6}^{-1/4},
\eeqa
a photosphere appears in the coasting phase and almost all the energy goes 
into the kinetic luminosity of the outflow $L_{\rm kin} \sim L_{0}$,
while a photosphere appears in the acceleration phase if $\eta > \eta_{*}$.
We can derive
\beqa
\Gamma_{f}&=&\min[\eta,\eta_{*}],
\quad
r_{\rm ph}/r_{0}=\max[\eta_{*}(\eta/\eta_{*})^{-3},\eta_{*}(\eta/\eta_{*})^{-1/3}],
\nonumber\\
L_{\rm kin}/L_{0}&=&\min[1,\eta_{*}/\eta],
\quad
L_{\rm ph}/L_{0}=\min[(\eta/\eta_{*})^{8/3},1],
\quad
T_{\rm ph}/T_{0}=\min[(\eta/\eta_{*})^{8/3},1],
\label{eq:fire}
\eeqa
where the first (second) value in the bracket
is for $\eta<\eta_{*}$ ($\eta>\eta_{*}$),
$\Gamma_{f}$ is the final Lorentz factor,
a thermal photosphere at radius $r_{\rm ph}$ has
a luminosity $L_{\rm ph}$ with an observed temperature $T_{\rm ph}$,
and we neglect finite shell effects for simplicity \citep{meszaros02}.
The above relations hold provided the fireball is not too baryon-poor,
$\eta > \eta_{\pm} \sim 10^{5} L_{47.5}^{1/4} r_{0,6}^{1/2}$ \citep{meszaros00}.

Internal shocks in a variable outflow with 
$\Gamma=10^{2}\Gamma_{2}$ are expected to occur at radii
\beqa
r_{s} \simeq 2 \Gamma^{2} c \Delta t \sim 6 \times 10^{13} 
\Gamma_{2}^{2} \Delta t_{-1} {\rm \ cm}
\eeqa
where $\Delta t=10^{-1}\Delta t_{-1}$ s is the variability timescale.
The minimum value is $\Delta t \sim r_{0}/v_{A} \sim 3 \times 10^{-5}$ s
since the Alfven velocity in the magnetosphere is $v_{A} \sim c$,
which is consistent with the initial rise time $\siml 0.3$ ms
\citep{terasawa05,palmer05}.
If shocks occur above the photosphere, i.e., 
$\eta > \eta_{s}=10 L_{0,47.5}^{1/5} \Delta t_{-1}^{-1/5}$,
nonthermal radiation is produced.
The nonthermal shock luminosity $L_{s}=\xi_{s} L_{\rm kin}$ 
dominates the photospheric luminosity $L_{\rm ph}$
if $\eta < 100 \eta_{*,2} \xi_{s}^{3/8}$, where $\xi_{\rm s}$ is 
the conversion efficiency of kinetic energy into photons.

Two typical scenarios are possible for the December 27 flare.
The first one is a baryon-poor scenario, e.g. $\eta \sim 10^{4} > \eta_{*}$,
where the photosphere is in the acceleration phase.
Most of the energy is radiated as photospheric emission, which may 
explain the thermal spectrum with temperature 
$T_{\gamma} \sim 175\pm 25$ keV observed by \citet{hurley05}.
The remaining kinetic energy is $\eta/\eta_{*} \sim 10^2$
times smaller than the radiation energy, which is also
implied by the radio afterglow of SGR 1806-20,
if we use typical parameters inferred from GRB afterglow fittings
$\xi_{e} \sim 0.1$, $\xi_{B} \sim 0.01$ and $n \sim 1$ cm$^{-3}$
\citep{nakar05,wang05}.

The second is a baryon-rich scenario, e.g. with $\eta \sim 10 < \eta_{*}$.
The observed thermal radiation $L_{\gamma}$ can be explained by
a thermalized shock luminosity $L_{s}=\xi_{s} L_{\rm kin}$, if
shocks occurred right below the photosphere \citep{rees04}.
The remaining kinetic energy is
$(1-\xi_{s})/\xi_{s} \sim 10 \xi_{s,-1}^{-1}$ times
larger than the radiation energy,
so that the observed radio afterglow may require
a jet configuration ($\theta \lesssim 0.1$) or atypical model parameters
(high ambient density $n$ and low $\xi_{e}$ or $\xi_{B}$).
However a jet may be implied by the elliptical image
and polarization of the radio afterglow \citep{gaensler05}
as well as the light curve of the giant flare that is
well fitted by emission from a relativistic jet \citep{yamazaki05}.
Atypical parameters may be also suggested by 
the rapid decay of the radio afterglow \citep{cameron05}.
Thus this model may also be viable.

The actual spectrum of the November 27 flare, however, may be nonthermal 
\citep{palmer05,mazets05}.
At least a portion of the giant flare may be nonthermal, 
since \citet{hurley05} determined the spectrum with low time resolution.
A previous giant flare in SGR 0526-66 may also have had
a nonthermal spectrum \citep{fenimore96}.
If giant flares are observed as short GRBs,
their spectra are also likely nonthermal \citep{fenimore96,nakar05b}.
A nonthermal flare component may arise in a baryon-rich model $\eta \simg 10$, 
since internal shocks extending above the photosphere naturally produce 
nonthermal emission.  
The characteristic synchrotron frequency 
$\epsilon_{m}=\Gamma \gamma_{m}^2 \hbar q B/m_{e} c$ in internal shocks
can be estimated as
\beqa
\epsilon_{m} \sim 300 \xi_{B}^{1/2} \xi_{e}^{3/2}
L_{\gamma,47.5}^{1/2} \Gamma_{1}^{-2} \Delta t_{-1}^{-1}
{\rm \ keV},
\eeqa
where a fraction $\xi_{e}$ of the internal energy goes
into electrons ($\gamma_{m} \sim \xi_{e} m_{p}/m_{e}$)
and a fraction $\xi_{B}$ goes into the magnetic field, 
$4 \pi r_{s}^{2} c \Gamma^{2} B^{2}/8\pi = 
\xi_{B} L_{\gamma}/\xi_{e}$.
Note that this synchrotron emission is not thermalized above the photosphere.

\section{Proton interactions and neutrinos}\label{sec:neutrino}

According to \S~\ref{sec:fire}, we choose the particular
parameters for the two typical models as $${\rm baryon-poor
~(BP):} ~\eta \sim 10^4, ~L_{\rm kin} \sim 10^{-2} L_{\gamma} \sim
10^{45.5} ~{\rm erg ~s}^{-1}, ~\Gamma \sim 100, ~\Delta t \sim 10^{-4}
~{\rm s},$$ $${\rm baryon-rich ~(BR):} ~\eta
\sim 10, ~L_{\rm kin} \sim 10 L_{\gamma} \sim 10^{48.5} ~{\rm erg
~s}^{-1}, ~\Gamma \sim 10, ~\Delta t \sim 10^{-1} ~{\rm s}.$$ 
We set $\Delta t$ in the baryon-poor model 
to maximize neutrino events, which turn out to be undetectable.
In the baryon-poor (baryon-rich) model internal shocks take place outside
(inside) the photospheric radius ($\tau_{\rm Th} \sim 1$).
Since the internal shocks are mildly relativistic, protons are
expected to be accelerated to a power law distribution,
$dn_p/d\epsilon_p \propto \epsilon_p^{-2}$
\citep{waxman97,waxman95,vietri95}. High energy protons may then
interact with photons ($p\gamma$) or with other cold protons ($pp$) to
produce neutrinos mostly through pion decays.  
The maximum shocked proton energy in the comoving equipartition magnetic field:
$B' = \sqrt{8\pi n'_p m_p c^2 \xi_B \xi_i}$, where $n'_p$ is the
comoving baryon number density, is limited by the system size and
synchrotron losses. By equating the acceleration time $t_{\rm acc}'
\sim \epsilon_p'/c q B'$ to the shorter of the comoving time $t_{\rm
com}' \sim r_s/\Gamma c$ and the synchrotron cooling time $t_{\rm
syn}' \sim 6 \pi m_p^4 c^3/\sigma_T m_e^2 \epsilon_p' B^{'2}$ we find
the maximum proton energy in the lab-frame as
\beqa 
\epsilon_{p,\rm max} \sim \cases{4\times 10^{16}~
(\xi_{B,-2} \xi_{i,-1} L_{\rm kin, 45.5})^{1/2} \Gamma_2^{-1} {\rm \
eV} & (BP) \cr 7\times 10^{18}~ (\xi_{B,-2} \xi_{i,-1} L_{{\rm
kin},48.5})^{-1/4} \Gamma_1^{5/2} \Delta t_{-1}^{1/2} {\rm \ eV} & (BR), }
\label{eq:p-max} 
\eeqa
where a fraction $\xi_{i} \sim 0.1 \xi_{i,-1}$ of the kinetic
energy goes into the internal energy. Note that $t_{\rm com}'$ is
smaller (greater) than $t_{\rm syn}'$ in the baryon-poor (baryon rich) model.

The comoving baryon density in the two models is
\beqa
n'_p = \frac{L_{\rm kin}}{4\pi r_s^2 \Gamma^2 m_p c^3} \sim 
\cases{ 2
\times 10^{11} L_{\rm kin, 45.5} \Gamma_2^{-6} \Delta t_{-4}^{-2} 
~{\rm cm}^{-3} & (BP) \cr 2
\times 10^{14} L_{\rm kin, 48.5} \Gamma_1^{-6} \Delta t_{-1}^{-2} 
~{\rm cm}^{-3} & (BR). }
\label{eq:p-dens}
\eeqa
The accelerated proton flux that would be measured at Earth, if they were to reach
us in a straight line before converting to neutrinos is
\beqa
\Phi_p = \frac{\xi_i L_{\rm kin}}{4 \pi d^2 \epsilon_p^2} &\sim & 
\cases{ 20 ~(\epsilon_{p}/{\rm GeV})^{-2} \xi_{i,-1} L_{\rm kin, 45.5} 
d_{1}^{-2} ~{\rm GeV}^{-1} ~{\rm cm}^{-2} ~{\rm s}^{-1} & (BP) 
\cr 2 \times 10^4 ~(\epsilon_{p}/{\rm GeV})^{-2}
\xi_{i,-1} L_{\rm kin, 48.5} d_{1}^{-2}
~{\rm GeV}^{-1} ~{\rm cm}^{-2} ~{\rm s}^{-1} & (BR). } 
\label{eq:p-flux}
\eeqa
A fraction of this proton flux will be converted to neutrinos
depending on the opacity of $p\gamma$ and $pp$ interactions in the
fireball. 

The photospheric thermal radiation bathes the ejecta at the internal
shock region (which also radiates), so that accelerated protons
interact with photons (using the observed $T_\gamma$) of comoving
energy 
\beqa
\epsilon_{\gamma}' \sim T_{\gamma}/\Gamma \sim \cases{ 2~
\Gamma_{2}^{-1} ~{\rm keV} & (BP) \cr 20~ \Gamma_{1}^{-1} ~{\rm keV} & (BR) }
\eeqa 
and can produce pions if the observed proton energy is
\beqa
\epsilon_p \sim \frac{0.3 \Gamma^2 {\rm GeV}^{2}}
{\epsilon_{\gamma}} \sim \cases{ 2 \times 10^{16}
\Gamma_2^2 \epsilon_{\gamma,5.3}^{-1} ~{\rm eV} & (BP) \cr 
2 \times 10^{14} \Gamma_1^2 \epsilon_{\gamma,5.3}^{-1} ~{\rm eV} &
(BR), }
\label{eq:bb-p-energy}
\eeqa
which is below the maximum available proton energy in both the
models. The density of these thermal photons at the shocks is\footnote{
Below the photosphere the photon density could be larger than this estimate.}
\beqa
n_{\gamma}' \sim \frac{L_{\gamma}}{4\pi r_s^2 \Gamma^2 c
\epsilon_{\gamma}'}
\sim \cases{ 7 \times 10^{18} L_{\gamma,47.5} \epsilon_{\gamma,5.3}^{-1}
 \Gamma_{2}^{-5} \Delta t_{-4}^{-2} ~{\rm cm}^{-3} & (BP) \cr 7
 \times 10^{17} L_{\gamma,47.5} \epsilon_{\gamma,5.3}^{-1}
 \Gamma_{1}^{-5} \Delta t_{-1}^{-2} ~{\rm cm}^{-3} & (BR). }
\label{eq:bb-dens}
\eeqa
The corresponding optical depth to $p\gamma$ interactions is then
\beqa
\tau_{p\gamma} \sim \frac{\sigma_{p\gamma} n_{\gamma}' r_s} {\Gamma}
\sim \cases{ 2 L_{\gamma,47.5} \epsilon_{\gamma,5.3}^{-1} 
\Gamma_{2}^{-4} 
\Delta t_{-4}^{-1} & (BP) \cr 20 L_{\gamma,47.5} 
\epsilon_{\gamma,5.3}^{-1} \Gamma_{1}^{-4} 
\Delta t_{-1}^{-1} & (BR), }
\label{eq:opt-bb}
\eeqa
where $\sigma_{p\gamma} \sim 5 \times 10^{-28}$ cm$^2$ is the cross-section 
at the $\Delta$ resonance.

Protons lose $\sim 20\%$ of their energy at each $p\gamma$
interaction, dominated by the $\Delta$ resonance.  Approximately half
of the pions are charged and decay into high energy neutrinos $\pi^+
\to \mu^+ + \nu_{\mu} \to e^+ + \nu_e + \bar \nu_{\mu} + \nu_{\mu}$,
with the energy distributed roughly equally among the decay products.
Thus the neutrino energy is $\sim 5\%$ of the proton energy. From
equation (\ref{eq:bb-p-energy}), we find the typical neutrino energy
expected from the $p\gamma$ interactions with thermal photons as
\beqa
\epsilon_{\nu} \sim \frac{0.3 \Gamma^2 {\rm GeV}^{2}}
{20 \epsilon_{\gamma}} \sim \cases{ 8 \times 10^{5}
\Gamma_2^2 \epsilon_{\gamma,5.3}^{-1} ~{\rm GeV} & (BP) 
\cr 8 \times 10^{3} \Gamma_1^2 \epsilon_{\gamma,5.3}^{-1} ~{\rm
GeV} & (BR). }
\label{eq:bb-nu-energy}
\eeqa
The corresponding monoenergetic neutrino flux at Earth 
(equal for $\nu_{\mu}$, $\nu_{\tau}$ and
$\nu_{e}$ after oscillations in vacuum where ${\bar \nu}_{\mu}$
created from $\pi^+$ decay is transformed to $\nu_{\tau}$) 
can be found from equation (\ref{eq:p-flux}) as
\beqa
\Phi_{\nu, p\gamma} &=& {\rm min}(1, \tau_{p\gamma})
\frac{0.2}{8} \frac{\xi_i L_{\rm kin}}{4 \pi d^2 \epsilon_{\nu}^2}
\nonumber \\
&\sim & \cases{ 7 \times 10^{-13}~\xi_{i,-1} L_{\rm kin, 45.5}
\epsilon_{\gamma,5.3}^2
d_{1}^{-2} \Gamma_2^{-4} ~{\rm GeV}^{-1} ~{\rm cm}^{-2} ~{\rm s}^{-1} & (BP)
\cr 7 \times 10^{-6} ~\xi_{i,-1} L_{\rm kin, 48.5} \epsilon_{\gamma,5.3}^2
d_{1}^{-2} \Gamma_1^{-4} ~{\rm GeV}^{-1} ~{\rm cm}^{-2} ~{\rm s}^{-1} & (BR), }
\label{eq:bb-nu-flux}
\eeqa
where in both the cases ${\rm min}(1, \tau_{p\gamma}) = 1$
from equation (\ref{eq:opt-bb}).
Note that the synchrotron and inverse Compton losses of $\pi^{+}$ and $\mu^{+}$
are negligible for our models.

In addition to $p\gamma$ interactions,
shock accelerated protons may also undergo $pp$ interactions with cold
protons in the ejecta and produce $\pi^{\pm}$.
The opacity for $pp$ interactions is about $\tau_{pp} 
\sim \sigma_{pp} n_{p}' r_{s}/\Gamma \sim 0.6
~L_{\rm kin, 48.5} \Gamma_1^{-5} \Delta t_{-1}^{-1} \sim 0.1
\tau_{\rm Th}$ for an average $pp$ cross-section of $\sigma_{pp} \sim 6 \times 
10^{-26}$ cm$^2$ in the TeV-PeV energy range. We calculate the
neutrino flux from $\pi^{\pm}$ decays as \citep{razzaque03}
\beqa
\Phi_{\nu, pp} = {\rm min} (1, \tau_{pp}) \int^{\epsilon_{p,\rm max}} 
\Phi_p M_{\nu} (\epsilon_p) d\epsilon_p,
\eeqa
where the neutrino multiplicity from $pp$ interactions in units of GeV$^{-1}$
is given by
\beqa
M_{\nu}(\epsilon_p) =
\frac{7}{4} \left( \frac{\epsilon_{\nu}}{\rm
GeV} \right)^{-1} \left[ \frac{1}{2} {\rm ln}\left( \frac{10^{11}\,
{\rm GeV}}{\epsilon_p}
\right) \right]^{-1} \Theta \left( \frac{1}{4} \frac{m_{\pi}}{\rm GeV}
\gamma_{\rm cm} \leq \frac{\epsilon_{\nu}}{\rm GeV} \leq 
\frac{1}{4} \frac{\epsilon_p}{\rm GeV} \right),
\eeqa
$\gamma_{\rm cm}$ is the Lorentz factor of the $pp$ center of mass
in the lab-frame and $\Theta$ is a step function.
Note that the flux is given for muon (anti
muon) neutrinos. Electron and tau (and their anti) neutrino fluxes
would be the same on Earth. 
We may fit the $pp$ muon neutrino flux for $\tau_{pp}<1$ by
\beqa
\Phi_{\nu, pp} = \left[ 330+ 30 ~{\rm ln}\left( \frac{\epsilon_{\nu}} 
{\rm GeV} \right) \right] \left( \frac{\epsilon_{\nu}}{\rm GeV} \right)^{-2}
~\xi_{i,-1}L_{\rm kin, 48.5}^2 d_{1}^{-2} \Gamma_1^{-5} \Delta
t_{-1}^{-1} 
 ~{\rm GeV}^{-1} {\rm cm}^{-2} {\rm s}^{-1}.
\label{eq:pp-nu-flux}
\eeqa

For the baryon-poor model with optically thin internal shocks, an
additional neutrino component may arise due to $p\gamma$ interactions
with non-thermal synchrotron photons, but this latter flux component
is undetectably low and is ignored. Next we calculate the expected neutrino 
events at Earth.

\section{Neutrino events in AMANDA}\label{sec:dis}

SGR1806-20 is located in the southern sky at a declination of
$-20^{\circ}$. We have calculated the probability to detect muon
neutrinos with the AMANDA detector, located at the South pole at a depth of
1 km, by using a code which propagates neutrinos through Earth and
calculates the interaction rate in ice near the vicinity of the detector
\citep{razzaque04}. The resulting probability may be fitted in the 
TeV-EeV energy range with a broken power-law as
\beqa
P(\epsilon_{\nu}) = 7 \times 10^{-5} (\epsilon_{\nu}/10^{4.5} 
{\rm GeV})^{\beta},
\label{eq:det-prob}
\eeqa
where $\beta=1.35$ for $\epsilon_{\nu}<10^{4.5}$ GeV
while $\beta=0.55$ for $\epsilon_{\nu}>10^{4.5}$ GeV.
Using a geometrical detector area of $A_{\rm det} = 0.03$ km$^2$
and the flare duration $t_{0}=0.1$ s, 
the number of muon events from $p\gamma$ neutrinos ($\nu_{\mu}$) 
of energy given in equation (\ref{eq:bb-nu-energy}) is
\beqa
N_{\mu,p\gamma} &=& A_{\rm det} t_0 P(\epsilon_{\nu}) \epsilon_{\nu}
\Phi_{\nu, p\gamma} 
\nonumber\\
&\sim& \cases{0.007
~\min(1,\tau_{p\gamma}) \xi_{i,-1} L_{{\rm kin},45.5} d_{1}^{-2} 
(\epsilon_{\gamma,5.3}^{-1} \Gamma_2^2)^{\beta-1} A_{{\rm det},-1.5} t_{0,-1}
& (BP) \cr 20 
~\min(1,\tau_{p\gamma}) \xi_{i,-1} L_{{\rm kin},48.5} d_{1}^{-2} 
(\epsilon_{\gamma,5.3}^{-1} \Gamma_1^2)^{\beta-1} A_{{\rm det},-1.5} t_{0,-1}
& (BR). }
\label{eq:muon}
\eeqa
On the other hand, the number of muon events from $pp$ neutrinos would be
\beqa
N_{\mu, pp} &=& 2A_{\rm det} t_0 \int_{\rm TeV}^{\rm PeV} P(\epsilon_{\nu})
\Phi_{\nu,pp} d\epsilon_{\nu} 
\nonumber\\
&\sim& \cases{3 \times 10^{-6}
~\xi_{i,-1} L_{{\rm kin},45.5}^2 d_{1}^{-2} \Gamma_{2}^{-5} \Delta t_{-4}^{-1}
A_{{\rm det},-1.5} t_{0,-1}
& (BP) \cr 300 
~\xi_{i,-1} L_{{\rm kin},48.5}^2 d_{1}^{-2} \Gamma_{1}^{-5} \Delta t_{-1}^{-1}
A_{{\rm det},-1.5} t_{0,-1}
& (BR), }
\eeqa
where we use equation (\ref{eq:pp-nu-flux}) 
and note that Cherenkov detectors do not distinguish between $\nu_{\mu}$ and
${\bar \nu}_{\mu}$ flavors.
The difference of the arrival time between neutrinos and gamma-rays
would be within $\sim t_0 \sim 0.1$ s.
Neutrinos can precede gamma-rays if 
internal shocks occur deeply inside the photosphere.

\begin{figure}
\plotone{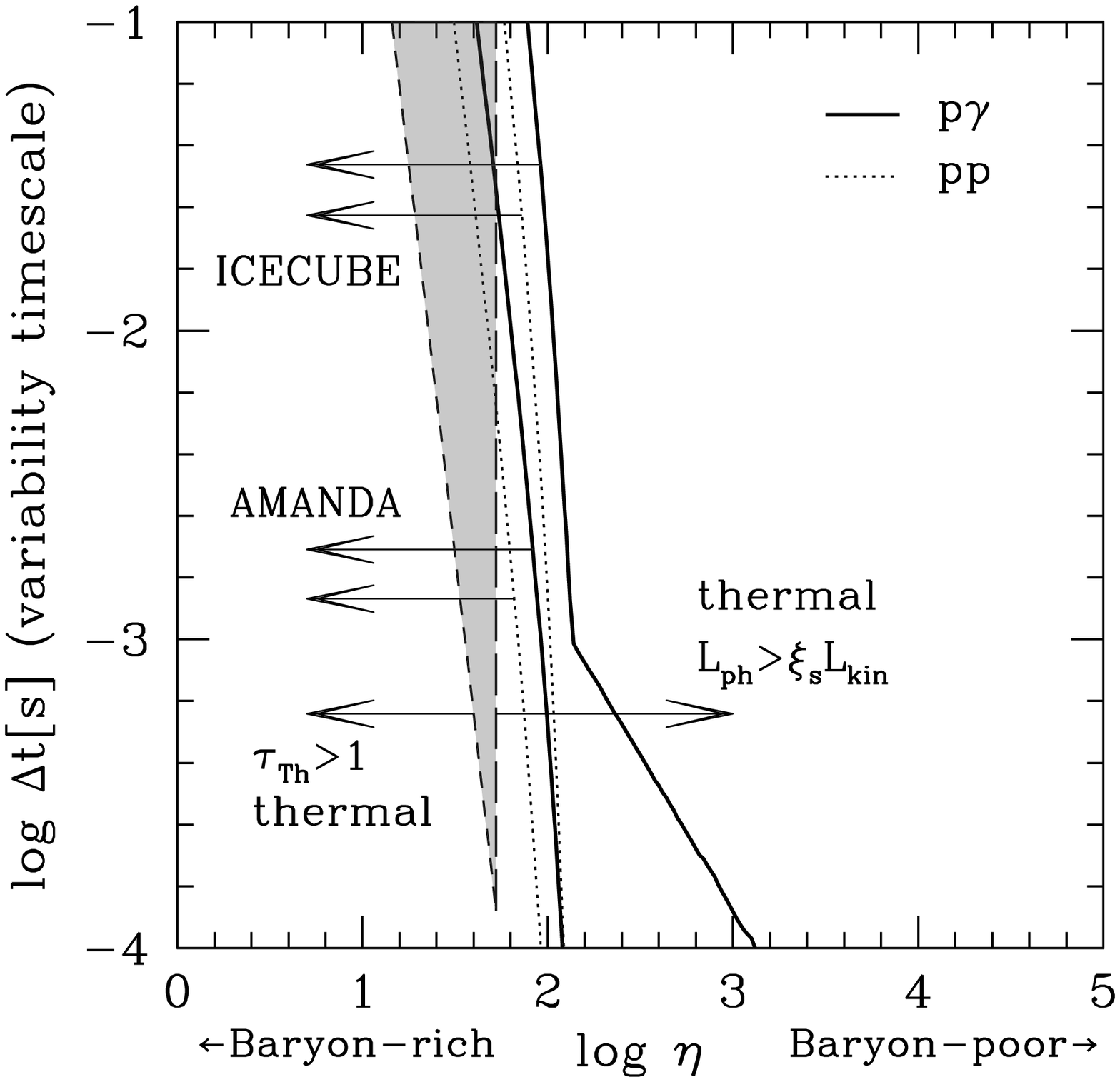}
\caption{\label{fig:para}
Parameter regions where AMANDA and ICECUBE 
can detect more than one muon event
are shown in the plane of the baryon load $\eta$ and 
the variability timescale $\Delta t$.
Baryons are rich (poor) for $\eta \siml 10^2$ ($\eta \simg 10^2$)
while the corresponding Lorentz factor is $\Gamma \sim \eta$ ($\Gamma \sim 10^2$)
in equation (\ref{eq:fire}).
Solid (dotted) lines are for $p\gamma$ ($pp$) neutrinos.
Photospheric thermal emission dominates the nonthermal emission 
$L_{\rm ph}>\xi_{s} L_{\rm kin}$ on the right of the long dashed line,
while internal shocks occur below the photosphere $\tau_{\rm Th}>1$ 
(also leading to a thermal spectrum) on the left of the dashed line. 
Thus the flare spectrum is nonthermal in the shaded region.
We used a normalization $\xi_{s} L_{\rm kin}+L_{\rm ph}=
L_{\gamma}=3 \times 10^{47}$ erg s$^{-1}$ and adopted
$\xi_{B}=0.01$, $\xi_{s}=0.1$ and $\xi_i=0.1$.
}
\end{figure}

In Figure~\ref{fig:para}, based on similar calculations, 
we show the parameter space where AMANDA and the future detector 
ICECUBE ($A_{\rm det}=1$ km$^{2}$) \citep{ahrens04}
can detect more than one muon event in the plane of
the baryon load $\eta$ and the variability timescale $\Delta t$.
(Note that background events are negligible.)
Baryons are rich (poor) for $\eta \siml 10^2$ ($\eta \simg 10^2$)
while the corresponding Lorentz factor is $\Gamma \sim \eta$ ($\Gamma \sim 10^2$)
in equation (\ref{eq:fire}).
We can see that TeV-PeV neutrinos may have been already detected by AMANDA
if the giant flare is baryon-rich ($\eta \siml 30$),
while a nondetection would suggest a baryon-poor fireball.
This offers the exciting prospect of gaining
independent information about the baryon load or the bulk Lorentz factor
of the fireball, the efficiency of proton injection and energy dissipation 
in shocks and the variability timescale associated with the flare trigger.
Such parameters would constrain the energetics and the radiation mechanisms 
inferred from electromagnetic observations.

If the neutrinos from this flare are detected by AMANDA, one 
would expect ICECUBE to be able to detect less energetic flares 
in this and other galactic SGRs. One can show that the baryon-rich 
model with a flare $\sim 10^{-3}$ times smaller than that considered 
here can produce about one event in ICECUBE, and the rate of such flares 
is about $\sim 1/10$ yr.

Since sedimentation due to gravity causes heavier elements to stratify down,
the surface tends to consist of lighter elements \citep{alcock80}. 
An absorption feature in bursts from SGR 1806-20 was interpreted as 
due to proton cyclotron lines \citep{ibrahim03,ho01}. However the magnetar 
surface may not contain hydrogen since hydrogen could burn very fast 
\citep{chang04}. 
If a fireball contains some heavy nuclei, photo disintegration processes 
with thermal photons may substantially reduce the maximum energy of the 
nuclei \citep{puget76}. A detailed calculation requiring a Monte Carlo
simulation is out of the scope of this Letter. Here we present
qualitative remarks if a nucleus of arbitrary mass number
$A$ and charge $Z$ survives photo disintegration while being accelerated in 
the internal shocks. Their maximum energy would be $\propto Z$ for
$t_{\rm com}'<t_{\rm syn}'$ and $\propto A^2 Z^{-3/2} \sim Z^{1/2}$ for
$t_{\rm com}'>t_{\rm syn}'$ in equation (\ref{eq:p-max}). 
The threshold energy of nuclei for the pion
production is $\sim A$ times equation (\ref{eq:bb-p-energy}). 
The pions decay into neutrinos and the typical neutrino
energy is almost the same as equation (\ref{eq:bb-nu-energy}). Since the
neutrino energy is the same, the neutrino flux is the same as equation
(\ref{eq:bb-nu-flux}). (Note that the optical depth for $\sim A$ pion
production by nuclei is also the same as equation (\ref{eq:opt-bb})
because the photomeson cross-section is $\sim A \tau_{p\gamma}$.)
Therefore the number of muon events in equation (\ref{eq:muon})
does not change much.

External shocks that produce the radio afterglows may also accompany
neutrino emissions. A simple application of the GRB afterglow
\citep{wang05}, however, shows that the typical synchrotron frequency is
too low to make the $p\gamma$ interactions.

Neutrons produced by $p\gamma$ interactions can reach us in a straight line
without decay if their energy is larger than $\sim 10^{18}$ eV
\citep[e.g.,][]{ioka04},
and may be observed as coincident cosmic rays.
Since $\epsilon_p \sim 10^{18}$ eV protons interact with 
$\sim 40 \Gamma_{1}^{2} \epsilon_{p,18}^{-1}$ eV flare photons,
the $p\gamma$ optical depth is about 
$\tau_{p\gamma} \sim 8\times 10^{-7} L_{\gamma,47.5} \Delta t_{-1}^{-1}
\epsilon_{p,18}^{-2}$
if we assume a thermal spectrum.
Then the number of cosmic ray events 
in $A_{\rm det}\sim 10^{3}$ km$^{2}$ detectors 
such as AUGER \citep{abraham04} may be
$N_{n}\sim \epsilon_p \Phi_p \tau_{p\gamma} A_{\rm det} t_0
\sim 10 \epsilon_{p,18}^{-3} \xi_{i,-1} L_{{\rm kin},48.5}^2 d_{1}^{-2} 
\Delta t_{-1}^{-1}$.

Neutral pions produced by $p\gamma$ and $pp$ interactions decay 
into two gamma-rays with a flux and energy comparable to neutrinos.
These gamma-rays might be detected by Milagro 
\citep{atkins03} if the flare had occured in the northern sky and
if the gamma-rays escape the emission region without making pairs.

\acknowledgments
We are grateful to C.~Thompson for discussions, to the referee for
comments, and to TIARA-Tsinghua University for hospitality (PM).
This work was supported in part by the Eberly Research Funds of Penn State 
and by the Center for Gravitational Wave Physics under grants 
PHY-01-14375 (KI,SK), NSF AST 0307376 (SR,PM) and NAG5-13286 (PM)
and by NASA Swift Cycle 1 GI program (SK).

%
%

%
%

\newpage

\end{document}